\pdfoutput=1
\documentclass[conference]{IEEEtran}
\IEEEoverridecommandlockouts
\usepackage{cite}
\usepackage{amsmath,amssymb,amsfonts}
\usepackage{algorithmic}
\usepackage{graphicx}
\usepackage{textcomp}
\usepackage{comment}
\usepackage{xcolor}
\usepackage{balance}
\usepackage{cite}
\usepackage{pdfpages}
\def\BibTeX{{\rm B\kern-.05em{\sc i\kern-.025em b}\kern-.08em
    T\kern-.1667em\lower.7ex\hbox{E}\kern-.125emX}}
\begin{document}

\title{\vspace{-3mm}\textcolor{black}{Analysis of Intelligent Vehicular Relaying in Urban 5G+ Millimeter-Wave Cellular Deployments\vspace{-1mm}}
\thanks{$^1$A part of this work has been completed during the research visit of Vitaly Petrov to The University of Texas at Austin, USA in Fall 2018.}
}

\author{\IEEEauthorblockN{Vitaly Petrov$^{\ast,1}$, Dmitri Moltchanov$^\ast$, Sergey Andreev$^\ast$, and Robert W. Heath Jr.$^\dagger$}
\IEEEauthorblockA{$^\ast$Unit of Electrical Engineering, Tampere University, Finland\\
$^\dagger$Department of Electrical and Computer Engineering, The University of Texas at Austin, USA\\
\{vitaly.petrov, dmitri.moltchanov, sergey.andreev\}@tuni.fi, rheath@utexas.edu\vspace{-2mm}}
}

\maketitle

\begin{abstract}
\textcolor{black}
{
The capability of smarter networked devices to dynamically select appropriate radio connectivity options is especially important in the emerging millimeter-wave (mmWave) systems to mitigate abrupt link blockage in complex environments. To enrich the levels of diversity, mobile mmWave relays can be employed for improved connection reliability. These are considered by 3GPP for on-demand densification on top of the static mmWave infrastructure. However, performance dynamics of mobile mmWave relaying is not nearly well explored, especially in realistic conditions, such as urban vehicular scenarios. In this paper, we develop a mathematical framework for the performance evaluation of mmWave vehicular relaying in a typical street deployment. We analyze and compare alternative connectivity strategies by quantifying the performance gains made available to smart devices in the presence of mmWave relays. We identify situations where the use of mmWave vehicular relaying is particularly beneficial. Our methodology and results can support further standardization and deployment of mmWave relaying in more intelligent 5G+ ``all-mmWave'' cellular networks.
}
\end{abstract}

\section{Introduction}
\label{sec:idea}
\label{sec:intro}
Millimeter-wave (mmWave) communications is one of the key solutions introduced by the fifth-generation (5G) wireless networks. Adopted by 3GPP for New Radio (NR) technology, mmWave communications enable transmissions with the data rates considerably higher than those offered by 4G microwave solutions~\cite{heath_mmwave}. 
\textcolor{black}{In contrast, the coverage range of a mmWave access point (AP) is expected to be smaller than that offered by sub-$6$\,GHz cellular systems and will remain within a few hundred meters}~\cite{akdeniz_channel_model}. The highly directional mmWave transmissions are also susceptible to blockage -- occlusion of the signal path by buildings, vehicles, and even human bodies~\cite{haneda_mmwave_blockage}. Dense deployments of mmWave APs are a natural solution, but may incur capital and operating expenditures~\cite{heath_dense_mmwave}.

Millimeter-wave relays are an alternative to backhauled APs. Static mmWave relays can densify the network at lower costs than full-fledged APs without compromising its performance~\cite{relay_mmwave2,relay_editor}. The use of static mmWave relays has been ratified by 3GPP as part of 5G~NR~Rel.~15~\cite{3gpp_static_relay}. Currently, 3GPP continues to investigate this area by targeting a possible adoption of mobile mmWave relays \textcolor{black}{mounted on vehicles and drones} as part of NR Rel. 17 and beyond~\cite{3gpp_mobile_relay}. That work is currently at an early stage focused primarily on identifying the target setups, where the deployment of mobile mmWave relays is especially beneficial. For this purpose, a holistic methodology is required, which accounts for a realistic city deployment, features of vehicular and drone-carried relay operation, and complex propagation of mmWave signal.

In this paper, we develop a mathematical framework for the performance evaluation of a cellular network with mmWave APs, \textcolor{black}{intelligent} mmWave users, and mmWave vehicular relays. Our framework accounts for the specifics of a realistic urban (street) deployment, 3GPP-compatible mmWave signal propagation model \textcolor{black}{with blockage caused by humans and vehicles}, and alternative relaying strategies. We apply our framework to quantifying the realistic performance gains that mmWave vehicular relays may bring to an average user. We highlight the conditions where the use of mmWave vehicular relays leads to a more than two-fold increase in the  spectral efficiency. Our methodology and numerical results can be used to justify the system design choices for the mmWave vehicular relaying in \textcolor{black}{complex and dynamic} mmWave-based networks.

\begin{figure*}[!t]
    \centering
    \includegraphics[width=0.95\textwidth]{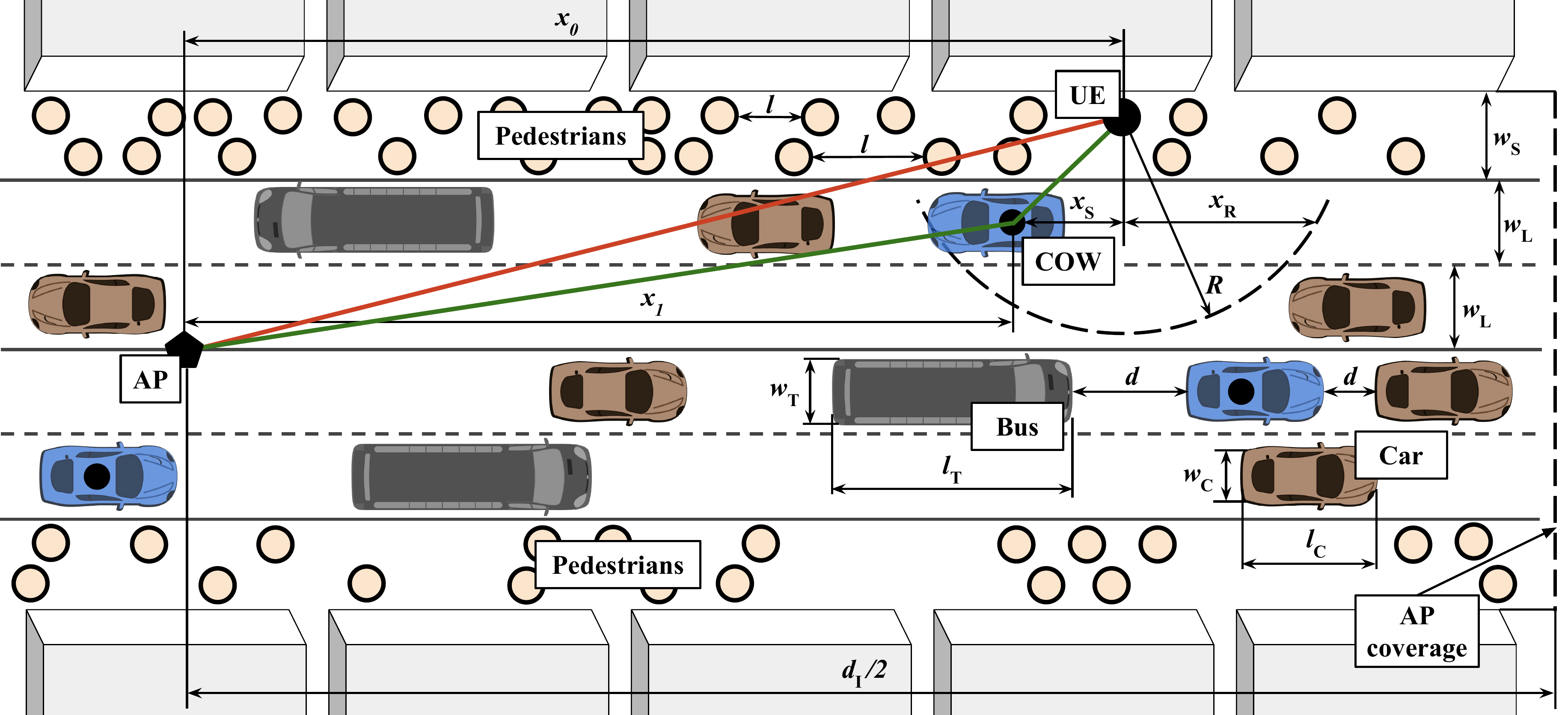}
    \vspace{-1mm}
    \caption{\textcolor{black}{Our considered urban street deployment for mmWave vehicular relaying with the regular placement of static mmWave APs, random locations of pedestrians, cars, and buses. A fraction $p_{\text{R}}$ of cars are also equipped with mmWave relaying capabilities and can forward traffic between UEs and mmWave~APs.}}
    \vspace{-3mm}
    \label{fig:street2}
\end{figure*}

\section{System Model}
\label{sec:system_model}

\subsection{Scenario and Deployment}
We consider a straight segment of a street with four traffic lanes and two sidewalks (see Fig.~\ref{fig:street2}). The lane width is $w_{\text{L}}$ and the width of the sidewalk is $w_{\text{S}}$. Static mmWave APs are deployed on the lampposts between the central lanes at a height $h_{\text{A}}$. The separation distance between the APs is $d_{\text{I}}$. On each sidewalk, there are two human paths representing a typical bidirectional flow. A human body is modeled as a cylinder with a radius $r_{\text{P}}$ and a height $h_{\text{P}}$. The pedestrians on the same path are separated by a random distance $\ell$, where $L$ is a generally-distributed random variable (RV) with the cumulative distribution function (CDF) $F_{L}(x)$. Each pedestrian carries a mmWave user equipment (UE) at a height of $h_{\text{U}}$.

Two types of vehicles are deployed in our scenario: (i)~regular vehicles termed \emph{cars} and modeled as parallelepipeds with the dimensions of $\ell_{\text{C}}$$\times$$w_{\text{C}}$$\times$$h_{\text{C}}$, and (ii)~large vehicles representing city buses and trucks termed \emph{buses} and modeled as parallelepipeds with the dimensions of $\ell_{\text{T}}$$\times$$w_{\text{T}}$$\times$$h_{\text{T}}$. Buses and cars are deployed randomly along the centers of all the traffic lanes with a random distance $d$ between their bumpers, where $D$ is a generally-distributed RV with the mean $\text{E}[D]$. Each vehicle is a bus with probability $p_{\text{T}}$ independently of other vehicles. \textcolor{black}{A fraction of cars, $p_{\text{R}}$,} are also equipped with mmWave relaying capabilities and can act as ``cells on wheels'' (COWs). The COW coverage range is $R$. 

\vspace{-1mm}
\subsection{Propagation Model}
\vspace{-1mm}
The mmWave signal propagation is modeled following the recent 3GPP considerations~\cite{3gpp_tr_38_901} and accounts for both human- and vehicle-body blockage. In case where the line-of-sight (LoS) path between the communicating nodes is occluded by either a human or a vehicle body, the nodes use an alternative non-line-of-sight (nLoS) path by utilizing one of the reflected or scattered mmWave signal components~\cite{gapeyenko_cluster_model}. 

Following~\cite{3gpp_tr_38_901}, the pathloss, $T$, is given by
\begin{align}\label{eq:pathloss}
T =
\begin{cases}
32.4+21.0\log_{10}(d_{\text{3D}})+20\log_{10}f_{\text{c}},\,\text{LoS},\\
32.4+31.9\log_{10}(d_{\text{3D}})+20\log_{10}f_{\text{c}},\,\text{nLoS},\\
\end{cases}
\end{align}
where $f_{c}$ is the carrier frequency and $d_{\text{3D}}$ is the 3D \textcolor{black}{separation} distance between the nodes.

The communicating entities in our model (APs, UEs, and COWs) set their transmit powers as $P_{\text{A}}$, $P_{\text{U}}$, and $P_{\text{C}}$, respectively. All the nodes also utilize directional antenna radiation patterns with the corresponding gains of $G_{\text{A}}$, $G_{\text{U}}$, and $G_{\text{C}}: G_{\text{U}} \leq G_{\text{C}} \leq G_{\text{A}}$. 
For simplicity, we assume perfect beam alignment between the communicating entities. 

\begin{figure*}[!t]
\vspace{-7mm}
\begin{align}\label{eq:both_pb}
p_{\text{B}}&=
\begin{cases}
F_{L}(z) + \frac{1 - F_{L}(z)}{\text{E}[L]} \left( 2z - \int^{2z}_{0} F_{L}(x)dx\right),\quad{}\quad{}\quad{}\quad{}\quad{}\quad{}\,\,\,\,\,\,h_{\text{T}} < h^{\star}_{\text{T}} \,\cap\, w_{\text{S}} \leq 2r_{\text{P}} + \frac{\left(h_{\text{P}} - h_{\text{U}}\right)\left( 8w_{\text{L}} + 3 w_{\text{S}}\right)}{2(h_{\text{A}} - h_{\text{U}})},\\
F_{L}\left(z\right),\quad{}\quad{}\quad{}\quad\quad{}\quad{}\quad{}\quad{}\quad{}\quad{}\quad{}\quad{}\quad{}\quad{}\quad{}\quad{}\quad{}\quad{}\quad{}\quad{}\,\,h_{\text{T}} < h^{\star}_{\text{T}} \,\cap\, w_{\text{S}} > 2r_{\text{P}} + \frac{\left(h_{\text{P}} - h_{\text{U}}\right)\left( 8w_{\text{L}} + 3 w_{\text{S}}\right)}{2(h_{\text{A}} - h_{\text{U}})},\\
1 - \frac{(\ell_{\text{C}}-p_{\text{T}}\ell_{\text{C}} + \text{E}[D])( 1 - F_{L}(z) - \frac{1 - F_{L}(z)}{\text{E}[L]}( 2z - \int^{2z}_{0} F_{L}(x)dx))}{p_{\text{T}}\ell_{\text{T}} + (1-p_{\text{T}})\ell_{\text{C}} + \text{E}[D]},\,\,\,\,\,\,\,\,\,\,\,h_{\text{T}} \geq h^{\star}_{\text{T}} \,\cap\, w_{\text{S}} \leq 2r_{\text{P}} + \frac{\left(h_{\text{P}} - h_{\text{U}}\right)\left( 8w_{\text{L}} + 3 w_{\text{S}}\right)}{2(h_{\text{A}} - h_{\text{U}})},\\
1 - \frac{\left(\ell_{\text{C}}-p_{\text{T}}\ell_{\text{C}} + \text{E}[D]\right)\left( 1 - F_{L}(z)\right)}{p_{\text{T}}\ell_{\text{T}} + (1-p_{\text{T}})\ell_{\text{C}} + \text{E}[D]},\quad{}\quad{}\quad{}\quad{}\quad{}\quad{}\quad{}\quad{}\quad{}\quad{}\quad{}\,\,\,\,\,h_{\text{T}} \geq h^{\star}_{\text{T}}\,\cap\, w_{\text{S}} > 2r_{\text{P}} + \frac{\left(h_{\text{P}} - h_{\text{U}}\right)\left( 8w_{\text{L}} + 3 w_{\text{S}}\right)}{2(h_{\text{A}} - h_{\text{U}})}.
\end{cases}
\end{align}
\vspace{-3mm}
\hrulefill
\normalsize
\end{figure*}

\subsection{Connectivity Models}
\label{sec:connectivity_models}

We analyze and compare three UE connectivity strategies:
\begin{itemize}

\item \textit{Baseline}. All the UEs always utilize the infrastructure link to the nearest static mmWave AP. No relays are used.

\item \textit{Conservative Relay}. COWs can assist UEs within their coverage. The radio resources occupied by UE-COW connections \textbf{do not overlap} with those utilized for UE-AP and COW-AP links. This strategy primarily reflects the implementation of mobile relays in 3G and 4G systems by providing a pessimistic estimate for the performance gains of mmWave vehicular relays in our scenario.

\item \textit{Agressive Relay}. COWs can assist UEs within their coverage. The radio resources occupied by UE-COW connections \textbf{may overlap} with those utilized for UE-AP and COW-AP links, thus exploiting better spatial diversity of narrow-beam 5G mmWave communications~\cite{petrov_twc}. This strategy offers an optimistic estimate for the performance gains of mmWave vehicular relays in our scenario.
\end{itemize}

For both relay-aided strategies, each of the UEs continiously selects the path currently characterized by the highest signal-to-noise ratio (SNR) out of those provided by the static APs and COWs. UE is assumed to instantaneously switch to the best available link via multi-connectivity mechanisms~\cite{rangan_mc}. 

In the following sections, we develop a mathematical framework for evaluating the introduced connectivity strategies.
We particularly focus on a dense deployment of mmWave APs that do not permit outages. Therefore, the UE mean spectral efficiency (SE) is selected as a primary performance indicator.

\section{Analysis of Baseline Model}\label{sec:direct_analysis}

\subsection{Human-Body Blockage Modeling}\label{sec:hbb}

\begin{figure}[!b]
    \centering
    \vspace{-5mm}
    \includegraphics[width=1.0\columnwidth]{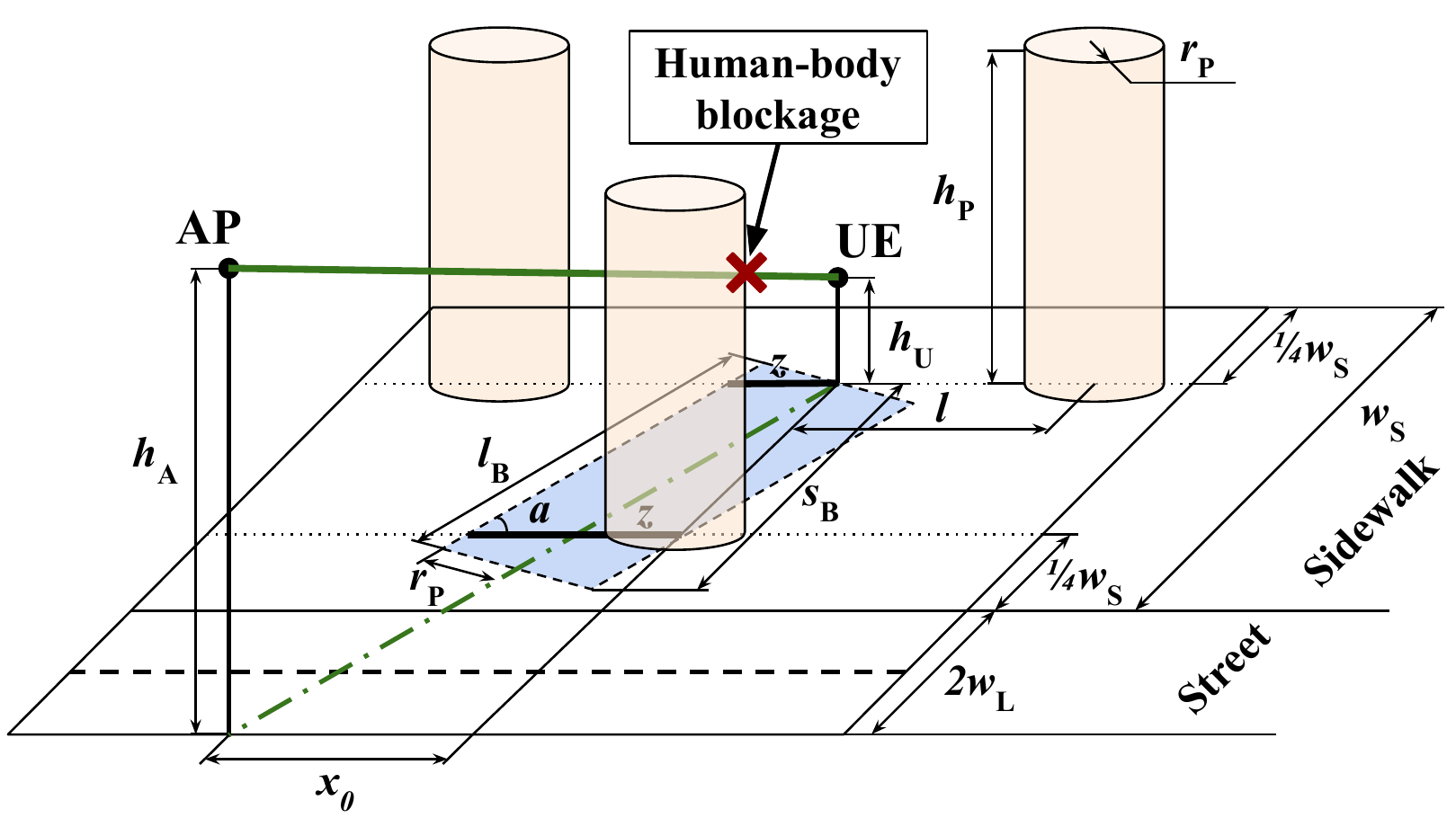}
    \vspace{-7mm}
    \caption{Human-body blockage modeling.}
    \vspace{-2mm}
    \label{fig:human_blockage}
\end{figure}

In this subsection, we derive the probability that the link between UE and AP is blocked by a pedestrian, where the UE is separated from the AP by a fixed distance of $x_{0}$. Following~\cite{petrov_jsac_mc}, the link is considered blocked if there is at least a single pedestrian center in the ``blockage zone'', see Fig.~\ref{fig:human_blockage}. The width of this rectangle is $2r_{\text{P}}$, while its length, $\ell_{\text{B}}$, can be derived as $\ell_{\text{B}} = r_{\text{P}} + d_{\text{2D}}(h_{\text{P}} - h_{\text{U}})/(h_{\text{A}}-h_{\text{U}})$,
where $d_{\text{2D}} = \sqrt{(3w_{\text{S}}/4 + 2w_{\text{L}})^2 + x^2_{0}}$ is the AP-UE distance.

Observe that the link can be blocked by the pedestrians on both paths. From the scenario geometry, the blockage probability for the pedestrians on the same path, $p_{\text{B}, \text{H}_{1}}$, equals the probability that at least a single cylinder base center is within the interval of length $z = r_{\text{P}} / \sin(\alpha)$. Hence, we have
\begin{align}
p_{\text{B}, \text{H}_{1}} = \textrm{Pr}\left \{ \ell \leq z \right \} = F_{L}(r_{\text{P}}/\sin\alpha),
\label{eq:human_ph1}
\end{align}
where $F_{L}(x)$ is the CDF of the RV characterizing the distance between the neighboring humans on the same path.

The link can also be blocked by the pedestrians on the other path when the following two conditions apply simultaneously. First, $\ell_{\text{B}}$ has to be large enough so that this path crosses the blockage zone rectangle. Second, there should be at least one cylinder base center within the $2z$-long segment of the path crossing the blockage zone rectangle (see Fig.~\ref{fig:human_blockage}).

The first condition can be written as $w_{\text{S}}/2 \leq s_{\text{B}}$, where $w_{\text{H}} = 2w_{\text{L}} + 3w_{\text{S}}/4$. For the second condition, we apply the result from~\cite{cox}. Finally, the sought blockage probability is
\begin{align}
p_{\text{B}, \text{H}_{2}}=
\begin{cases}
\frac{2z - \int^{2z}_{0} F_{L}(x)dx}{\text{E}[L]},\, w_{\text{S}} \leq 2r_{\text{P}} + 2w_{\text{H}}\frac{h_{\text{P}} - h_{\text{U}}}{h_{\text{A}} - h_{\text{U}}},\\
0,\quad{}\quad{}\quad{}\quad{}\quad{}\,\,\, w_{\text{S}} > 2r_{\text{P}} + 2w_{\text{H}}\frac{h_{\text{P}} - h_{\text{U}}}{h_{\text{A}} - h_{\text{U}}}.
\end{cases}
\end{align}

Finally, because the events of blockage for the two paths are independent of each other, we arrive at the following expression for the human-body blockage probability
\begin{align}\label{eq:human_pb}
p_{\text{B}, \text{H}}\hspace{-1mm}&=\hspace{-1mm}
\begin{cases}
F_{L}(z)\hspace{-1mm}+\hspace{-1mm}\frac{(1 - F_{L}(z))\left( 2z - \int^{2z}_{0} F_{L}(x)dx\right)}{\text{E}[L]},\, w_{\text{S}} \leq w_{\text{U}},\\
F_{L}\left(z\right),\quad{}\quad{}\quad{}\quad{}\quad{}\quad{}\quad{}\quad{}\quad{}\quad{}\,\,\,\,\,\,w_{\text{S}} \leq w_{\text{U}},\\
\end{cases}
\end{align}
where $z = r_{\text{P}}\sqrt{(8w_{\text{L}} + 3w_{\text{S}})^2 + 16x^2_{0}}/\left(8w_{\text{L}} + 3w_{\text{S}}\right)$ and $w_{\text{U}}=2r_{\text{P}} + w_{\text{H}}\left(h_{\text{P}} - h_{\text{U}}\right)/(h_{\text{A}} - h_{\text{U}})$.

\begin{figure}[!t]
    \centering
    \vspace{-3mm}
    \includegraphics[width=1.0\columnwidth]{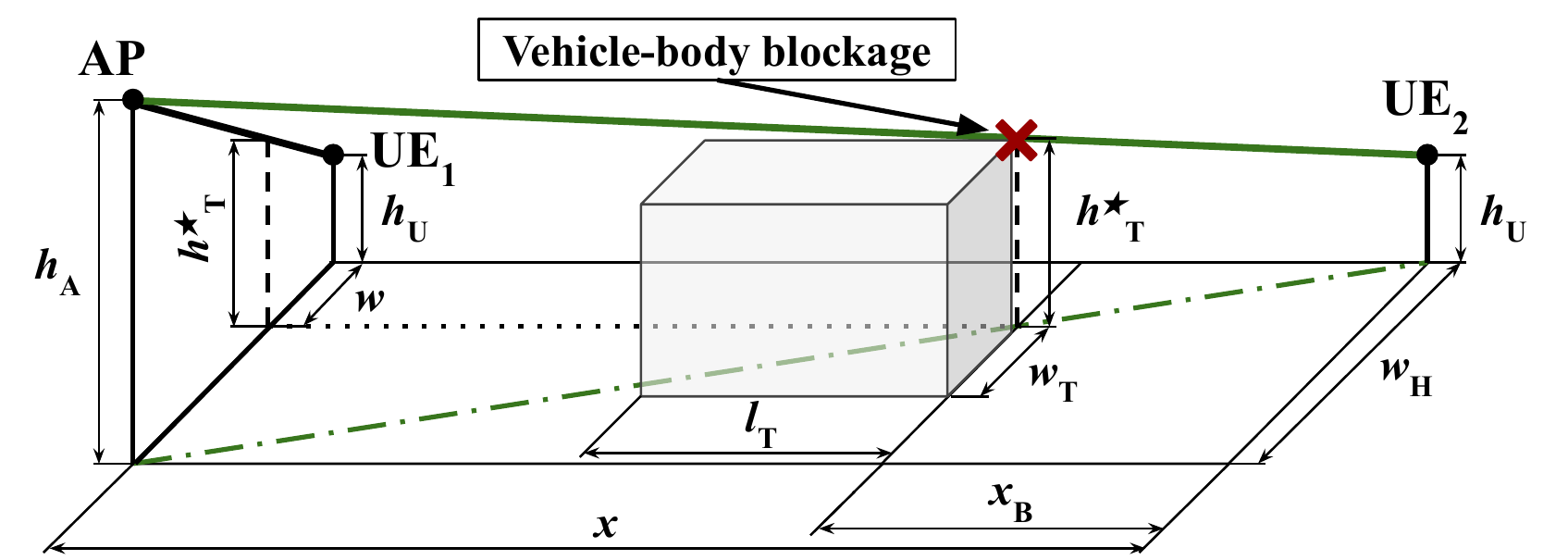}
    \vspace{-5mm}
    \caption{Minimal height of a bus resulting in vehicle-body blockage.}
    \vspace{-5mm}
    \label{fig:plot_lemma}
\end{figure}

\begin{figure*}[!t]
\vspace{-7mm}
\begin{align}\label{eq:p_blockage_cow_ap}
p^{\ast}_{\text{B}} &= 
\begin{cases}
F_{D_{\text{B}}}(\ell_{\text{B},\text{C}} - \frac{\ell_{\text{C}}}{2}),\quad{}\quad{} h_{\text{T}} < h^{\ast}_{\text{T}},\\
1 - (1 - \frac{p_{\text{T}}\ell_{\text{T}}}{p_{\text{T}}\ell_{\text{T}} + (1-p_{\text{T}})\ell_{\text{C}} + \text{E}[D]})(1 - F_{D_{\text{B}}}(\ell_{\text{B},\text{C}} - \frac{\ell_{\text{C}}}{2})),\quad{}\quad{} h_{\text{T}} \geq h^{\ast}_{\text{T}},\\
\end{cases}
\end{align}

\vspace{-2mm}
\begin{align}\label{eq:spectral_efficiency2_ue_cow_ap_x1}
\hspace{-5mm}C^{\dagger}_{2}(x_{0}, x_{\text{S}})\hspace{-1mm}&=\hspace{-1mm}\frac{1}{1 / C^{\star}(x_{\text{S}}) + 1 / C^{\ast}(x_{1})} =
p_{\text{C}} \Big(   \frac{p^{\star}_{\text{B}} p^{\ast}_{\text{B}}}{1/ \log_{2} (1 + S^{\star}_{\text{N}}) + 1 / \log_{2} (1 + S^{\ast}_{\text{N}})} +  \frac{(1 - p^{\star}_{\text{B}}) p^{\ast}_{\text{B}}}{1/ \log_{2} (1 + S^{\star}_{\text{L}}) + 1 / \log_{2} (1 + S^{\ast}_{\text{N}})} \nonumber\\ &+ \frac{p^{\star}_{\text{B}} (1 - p^{\ast}_{\text{B}})}{1/ \log_{2} (1 + S^{\star}_{\text{N}}) + 1 / \log_{2} (1 + S^{\ast}_{\text{L}})} + \frac{(1 - p^{\star}_{\text{B}})(1 - p^{\ast}_{\text{B}})}{1/ \log_{2} (1 + S^{\star}_{\text{L}}) + 1 / \log_{2} (1 + S^{\ast}_{\text{L}})  }     \Big).
\end{align}

\hrulefill
\vspace{-3mm}
\normalsize
\end{figure*}

\vspace{-1mm}
\subsection{Vehicle-Body Blockage Modeling}\label{sec:vbb}


In the considered scenario, a link between UE and AP can also be occluded by large vehicles, termed buses. Based on the scenario geometry in Fig.~\ref{fig:plot_lemma}, the following holds for the minimal bus height that results in blockage, $h^{\star}_{\text{T}}$
\begin{align}
\frac{h^{\star}_{\text{T}} - h_{\text{U}}}{h_{\text{A}} - h_{\text{U}}} = \frac{\sqrt{x^{2}_{\text{B}} + w^2}}{\sqrt{x^{2}_{0} + w^{2}_{\text{H}}}},
\label{eq:lemma_2}
\end{align}
where $x_{\text{B}}$ is the shift of the blocking vehicle from the UE, $x_{\text{B}} = x_{0}w/w_{\text{H}}$. Substituting $x_{\text{B}}$ into (\ref{eq:lemma_2}), we make an important observation that  $h^{\star}_{\text{T}}$ does not depend on $x_{0}$, i.e.,
\begin{align}
h^{\star}_{\text{T}} 
= h_{\text{U}} + \frac{3w_{\text{S}} + 2w_{\text{L}} - 2w_{\text{T}}}{8w_{\text{L}}+3w_{\text{S}}}(h_{\text{A}} - h_{\text{U}}).
\label{eq:lemma_4}
\end{align}



We now estimate the fraction of time when the UE-AP link is blocked by a bus. Since the number of cars between the two buses on the same lane follows the geometrical distribution with parameter $p_{\text{T}}$, the mean number of cars between the two buses, $\text{E}[N_{\text{C}}]$, can be estimated as $\text{E}[N_{\text{C}}] = (1 - p_{\text{T}})/p_{\text{T}}$. This implies that the random distance between the two buses, $d_{\text{B}}$, comprises of $N_{\text{C}}$ cars and $N_{\text{C}} + 1$ inter-vehicle intervals:
\begin{align}
\hspace{-1mm}\text{E}[D_{\text{B}}] = \text{E}[N_{\text{C}} \ell_{\text{C}} + (N_{\text{C}} + 1)d] = \frac{\text{E}[D] + \ell_{\text{C}} (1 - p_{\text{T}})}{p_{\text{T}}}.
\label{eq:vehicle_E_db}
\end{align}

Recalling that the deployment of vehicles follows the renewal process with generally distributed intervals, the vehicle-body blockage probability, $p_{\text{B}, \text{V}}$, can be established as~\cite{vencel_book_2}
\begin{align}\label{eq:vehicle_pb_intermediate}
p_{\text{B}, \text{V}}&=
\begin{cases}
0,\quad{}\quad{}\quad{}\,\,\,\,\,\,h_{\text{T}} < h^{\star}_{\text{T}},\\
\frac{\ell_{\text{T}}}{\ell_{\text{T}} + \text{E}[D_{\text{B}}]},\quad{}h_{\text{T}} \geq h^{\star}_{\text{T}}.\\
\end{cases}
\end{align}

\subsection{Mean Spectral Efficiency}\label{sec:baseline_se}


We now estimate the mean SE of the link between the UE and the AP. First, since the blockage events caused by humans and those caused by vehicles are independent, we derive the total blockage probability for the UE-AP link, $p_{\text{B}}$, as in (\ref{eq:both_pb}). Further, we determine the conditional SNR values in case of LoS (non-blocked) and nLoS (blocked) signal path between the UE and the AP, $S_{\text{L}}$ and $S_{\text{N}}$, respectively. Following (\ref{eq:pathloss}),
\begin{align}\label{eq:conditional_snr}
S_{\text{L}} &= 10^{\frac{P_{\text{U}} + G_{\text{A}} + G_{\text{U}} - N_{0}(B) - 32.4 - 20\log_{10}f_{\text{c}} - 21.0\log_{10}(K_{\text{U}})}{10}},\nonumber\\
S_{\text{N}} &= 10^{\frac{P_{\text{U}} + G_{\text{A}} + G_{\text{U}} - N_{0}(B) - 32.4 - 20\log_{10}f_{\text{c}} - 31.9\log_{10}(K_{\text{U}})}{10}},
\end{align}
where $K_{\text{U}} = \sqrt{x^2_{0} + [  2w_{\text{L}} + 3w_{\text{S}}/4 ]^2 + (h_{\text{A}}-h_{\text{U}})^2}$.

We obtain the mean SE for the UE located at a separation distance of $x_{0}$ from the AP, $C(x_{0})$, as
\begin{align}\label{eq:spectral_efficiency_baseline_x0}
C(x_{0}) = p_{\text{B}}\log_{2}(1 + S_{\text{N}}) + (1 - p_{\text{B}})\log_{2}(1 + S_{\text{L}}).
\end{align}

Finally, the mean SE for an arbitrary UE, $\text{E}[C]$, is derived~as
\begin{align}\label{eq:spectral_efficiency_baseline}
\text{E}[C] = \frac{2}{d_{\text{I}}}\int^{d_{\text{I}}/2}_{0} C(x_{0}) dx_{0}.
\end{align}

\section{Analysis of Relaying Models}
\label{sec:relay_analysis}
\subsection{UE-COW Link Analysis}

We now derive the mean SE for the relay link between the UE and the COW vehicle. We start by calculating the probability $p_{\text{C}}$ that there is at least a single COW within the interval $[x_{0} - x_{\text{R}}, x_{0} + x_{\text{R}}]$, where $x_{\text{R}}=\sqrt{ R^2-(3w_{\text{S}}/4 + w_{\text{L}}/2)^2 }$ is the maximum separation between the UE and the COW, so that the UE is under the COW coverage, $R$.

Recalling that only $p_{\text{R}}$ of cars act as COWs, we produce the mean distance between the neighboring COWs as $\text{E}[L_{\text{R}}] = \big[\ell_{\text{C}}(1-p_{\text{T}}) + \text{E}[D] + p_{\text{T}} \ell_{\text{T}}\big] / \big[p_{\text{R}}(1 - p_{\text{T}})\big]$.

Applying the approach from subsection~\ref{sec:vbb}, we obtain
\begin{align}
p_{\text{C}} = \left( 2x_{\text{R}} - \int^{2x_{\text{R}}}_{0} F_{L_{\text{R}}}(x)dx \right) / \text{E}[L_{\text{R}}].
\end{align}

Since we consider only the COWs deployed on the side lanes, the UE-COW link is not affected by vehicle-body blockage. Further, as the antenna array at the COW is assumed to be deployed on the rooftop of a car, the height of the COW, $h_{\text{C}}$, is considered to be lower than that of the UE, $h_{\text{U}}$. Therefore, the blockage zone rectangle (see subsection~\ref{sec:hbb} and Fig.~\ref{fig:human_blockage}) is always crossed by both human paths. Consequently, if there is a COW vehicle within the range of $R$ around the UE, the blockage probability for the UE-COW link, $p^{\star}_{\text{B}}$, can be directly obtained from (\ref{eq:human_pb}) as
\begin{align}\label{eq:p_blockage_ue_cow_xr}
\hspace{-3mm}p^{\star}_{\text{B}} = F_{L}(z_{1}) + \frac{1 - F_{L}(z_{1})}{\text{E}[L]} \left( 2z_{1} - \int^{2z_{1}}_{0} F_{L}(x)dx\right),
\end{align}
where $z_{1} = r_{\text{P}}\sqrt{(2w_{\text{L}} + 3w_{\text{S}})^2 + 16x^2_{s}}/(2w_{\text{L}} + 3w_{\text{S}})$, $x_{\text{S}}$ is a random separation distance between the COW and the UE~(see Fig.~\ref{fig:street2}): $x_{\text{S}} \in [-x_{\text{R}}, x_{\text{R}}]$.

\begin{figure}[!b]
    \centering
    \vspace{-3mm}
    \includegraphics[width=1.0\columnwidth]{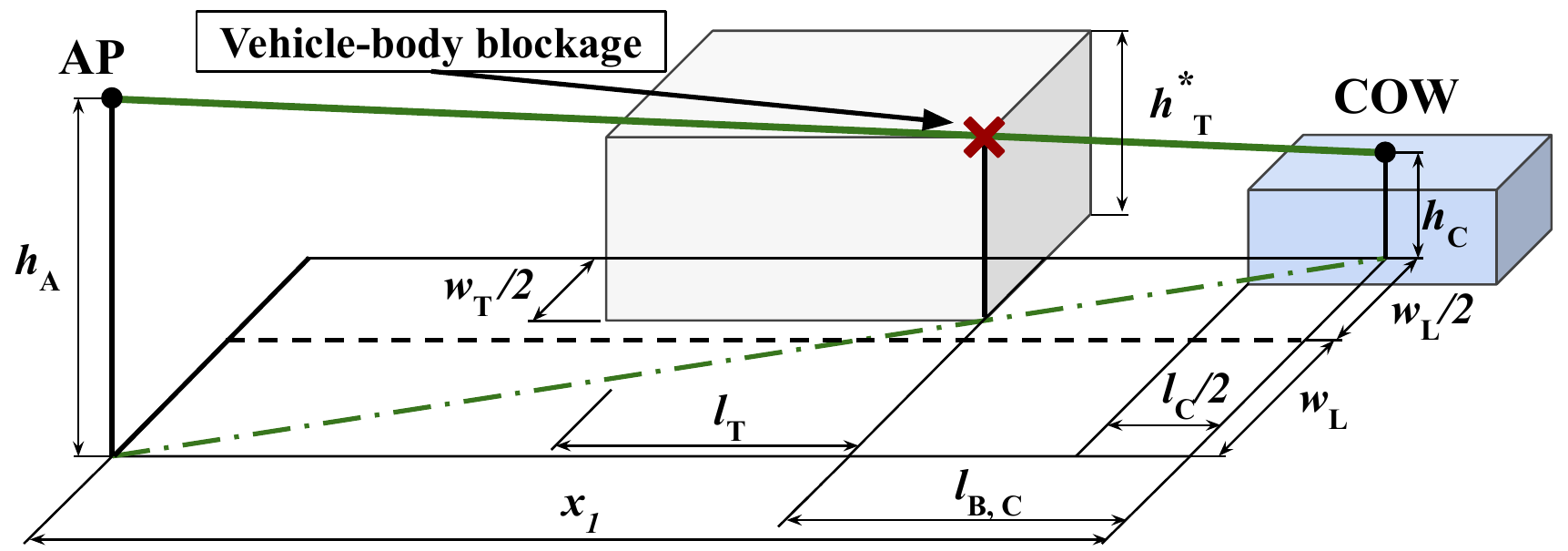}
    \vspace{-5mm}
    \caption{Minimal height of a bus leading to blockage of COW-AP link.}
    \label{fig:blockage_same_lane}
\end{figure}

Finally, the mean SE of the UE-COW link, $C^{\star}(x_{\text{S}})$ -- if there is at least a single COW in range and the UE selects a random COW out of those available -- can be computed as
\begin{align}
C^{\star}(x_{\text{S}}) &= p^{\star}_{\text{B}} \log_{2} \left( 1 + S^{\star}_{\text{N}} \right) + (1 - p^{\star}_{\text{B}}) \log_{2} \left( 1 + S^{\star}_{\text{L}} \right),
\end{align}
where the values $S^{\star}_{\text{L}}$ and $S^{\star}_{\text{N}}$ are obtained similarly to (\ref{eq:conditional_snr}).

\subsection{COW-AP Link Analysis}

Here, we obtain the mean SE of the link between the AP and the COW separated by $x_{1}$ from the AP. \textcolor{black}{This link is only affected by vehicle-body blockage from two sources}: (i) buses on the same lane and (ii) buses on the central lane.

For the latter case, the analysis is similar to that in subsection~\ref{sec:vbb}. Particularly, the blockage probability for the buses on the neighboring lane, $p^{\ast}_{\text{B}, \text{N}}$, can be written as
\begin{align}
p^{\ast}_{\text{B}, \text{N}} &=
\begin{cases}
0,\quad{}\quad{}\quad{}\,\,\,\,\, h_{\text{T}} < h^{\ast}_{\text{T}},\\
\frac{\ell_{\text{T}}}{\ell_{\text{T}} + \text{E}[D_{\text{B}}]},\quad{} h_{\text{T}} \geq h^{\ast}_{\text{T}},\\
\end{cases}
\label{eq:p_blockage_neighbor_lane}
\end{align}
where $h^{\ast}_{\text{T}} = h_{\text{C}} + (2w_{\text{L}} - w_{\text{T}})(h_{\text{A}} - h_{\text{C}})/3 w_{\text{L}}$ is the minimal height of a bus on the central lane yielding blockage.

Blockage of the COW-AP link by a bus on the same lane may occur if there is at least a single bus in the blockage zone of length $\ell_{\text{B},\text{C}}$, see Fig.~\ref{fig:blockage_same_lane}. For the given bus height and width, the blockage zone length, $\ell_{\text{B},\text{C}}$, has to be small enough to result in a blockage situation in both horizontal and vertical planes: $\ell_{\text{B},\text{C}} = x_{1} \min\big\{ \left(h_{\text{T}} - h_{\text{C}}\right) / \left(h_{\text{A}} - h_{\text{C}}\right) , w_{\text{T}} / 3w_{\text{L}}\big\}$,
where $x_{1} = x_{0} + x_{\text{s}}$.



The distance from an arbitrary COW vehicle to the nearest bus on the same lane is given by $d_{\text{B}} = N_{\text{C}}\ell_{\text{C}} + \sum^{N_{\text{C}}+1}_{i=1} d,$
where $N_{\text{C}}$ is a geometrically-distributed RV characterizing the number of cars between the COW and the nearest bus. The blockage probability by a bus on the same lane, $p^{\ast}_{\text{B}, \text{S}}$, is equal to the probability that $d_{\text{B}}$ does not exceed $\ell_{\text{B},\text{C}} - \ell_{\text{C}}/2$
\begin{align}
p^{\ast}_{\text{B}, \text{S}} = Pr\{ d_{\text{B}} \leq \ell_{\text{B},\text{C}} - \ell_{\text{C}}/2 \} = F_{D_{\text{B}}}(\ell_{\text{B},\text{C}} - \ell_{\text{C}}/2),
\label{eq:p_blockage_same_lane}
\end{align}
where $F_{D_{\text{B}}}$ is the CDF of the RV $D_{\text{B}}$ that can be computed numerically using non-linear RV transformation techniques~\cite{ross2014introduction}.

Observing that blockages by buses on the same and the neighboring lanes are independent events, joint blockage probability for the COW-AP link, $p^{\ast}_{\text{B}}$, is given in (\ref{eq:p_blockage_cow_ap}) as a combination of (\ref{eq:p_blockage_neighbor_lane}) and (\ref{eq:p_blockage_same_lane}). Finally, the mean SE for the COW-AP link where COW is separated from the AP by $x_{1}$~is
\begin{align}\label{eq:spectral_efficiency_cow_ap_x1}
C^{\ast}(x_{1}) &= p^{\ast}_{\text{B}}\log_{2}(1 + S^{\ast}_{\text{N}}) + (1 - p^{\ast}_{\text{B}})\log_{2}(1 + S^{\ast}_{\text{L}}).
\end{align}

The corresponding SNR values for the COW-AP link in LoS and nLoS cases, $S^{\ast}_{\text{L}}$ and $S^{\ast}_{\text{N}}$, are calculated similarly to (\ref{eq:conditional_snr}).

\subsection{Joint UE-COW-AP Connection Analysis}

In this subsection, we derive the mean SE for the relay-aided UE-COW-AP connection for both relaying strategies considered by our study as detailed in subsection~\ref{sec:connectivity_models}.


According to the \emph{Aggressive} strategy, the performance of the joint UE-COW-AP connection is limited exclusively by the mean SE of the COW-AP link. Therefore, the mean SE of the joint UE-COW-AP connection, $C^{\dagger}_{1}(x_{0}, x_{s})$, is equal to $C^{\ast}(x_{1})$. Its probability mass function (pmf), $f_{C^{\dagger}_{1}}(x)$, is thus
\begin{align}\label{eq:optimistic_se_pmf}
f_{C^{\dagger}_{1}}(x) = 
\begin{cases}
p_{\text{C}}p^{\ast}_{\text{B}},\quad{}\quad{}\,\,\,\,\,\,x = \log_{2}(1 + S^{\ast}_{\text{N}}),\\
p_{\text{C}}(1 - p^{\ast}_{\text{B}}),\,\,\,x = \log_{2}(1 + S^{\ast}_{\text{L}}),\\
1 - p_{\text{C}},\quad{}\,\,\,\,\,\,\,\,x = 0.\\
\end{cases}
\end{align}

With the \emph{Conservative} strategy, the radio resources allocated for UE-COW links \textcolor{black}{do} not overlap with those available for COW-AP and UE-AP links. Therefore, if the relay link UE-COW exists, the mean SE for the UE-COW-AP connection, $C^{\dagger}_{2}(x_{0}, x_{\text{S}})$, is obtained as in (\ref{eq:spectral_efficiency2_ue_cow_ap_x1}).

We now derive the mean SE when a smart UE selects the best available connection out of UE-AP and opportunistic UE-COW-AP. For this purpose, we first produce the pmf for the SE of the infrastructure UE-AP link (\emph{Baseline} strategy) where the UE is separated from the AP by $x_{0}$, $f_{C}(x)$. Recalling (\ref{eq:spectral_efficiency_baseline_x0}),
\begin{align}\label{eq:se_baseline_pmf}
f_{C}(x) =
\begin{cases}
1 - p_{\text{B}},\,\,\,x = \log_{2}(1 + S_{\text{L}}),\\
p_{\text{B}},\qquad{}\,x = \log_{2}(1 + S_{\text{N}}).\\
\end{cases}
\end{align}

Finally, we calculate the mean SE for the best connection with \emph{Aggressive} and \emph{Conservative} strategies ($C_{1}(x_{0}, x_{\text{S}})$ and $C_{2}(x_{0}, x_{\text{S}})$, respectively) as the maximum of two RVs representing the SE of infrastructure UE-AP and relay UE-COW-AP connections. Accordingly, the mean SE for an arbitrary UE is derived by numerical integration as
\begin{align}
\text{E}[C_{1}] =  \frac{1}{d_{\text{I}}x_{\text{R}}} \int^{d_{\text{I}}/2}_{0}   \int^{x_{0} + x_{\text{R}}}_{x_{0}-x_{\text{R}}} \big( C_{1}(x_{0}, x_{\text{S}}) dx_{\text{S}} \big) dx_{0},\nonumber\\
\text{E}[C_{2}] = \frac{1}{d_{\text{I}}x_{\text{R}}} \int^{d_{\text{I}}/2}_{0}   \int^{x_{0} + x_{\text{R}}}_{x_{0}-x_{\text{R}}} \big( C_{2}(x_{0}, x_{\text{S}}) dx_{\text{S}} \big) dx_{0}.
\end{align}

\section{Numerical Results}
\label{sec:numerical_results}


In this section, the obtained results are elaborated numerically. We model a street segment with mmWave APs operating at $28$\,GHz with $1$\,GHz of bandwidth and deployed $300$\,m apart from each other. Following the 3GPP considerations, the heights of APs and UEs are set to $10$\,m and $1.5$\,m, respectively~\cite{3gpp_tr_38_901}. The UE transmit power is $23$\,dBm and the antenna gains are given as $G_{\text{A}}=27$\,dB and $G_{\text{U}}=15$\,dB~\cite{mmwave_antenna_elements}. The radius of a human body is set to $0.3$\,m, while its \textcolor{black}{height is} $1.75$\,m. We assume $5$\% of large vehicles, i.e., $p_{\text{T}}=0.05$~\cite{large_vehicles_fraction}.

\subsubsection{The effect of human density}

\begin{figure}[!t]
    \centering
    \vspace{-3mm}
    \includegraphics[width=0.9\columnwidth]{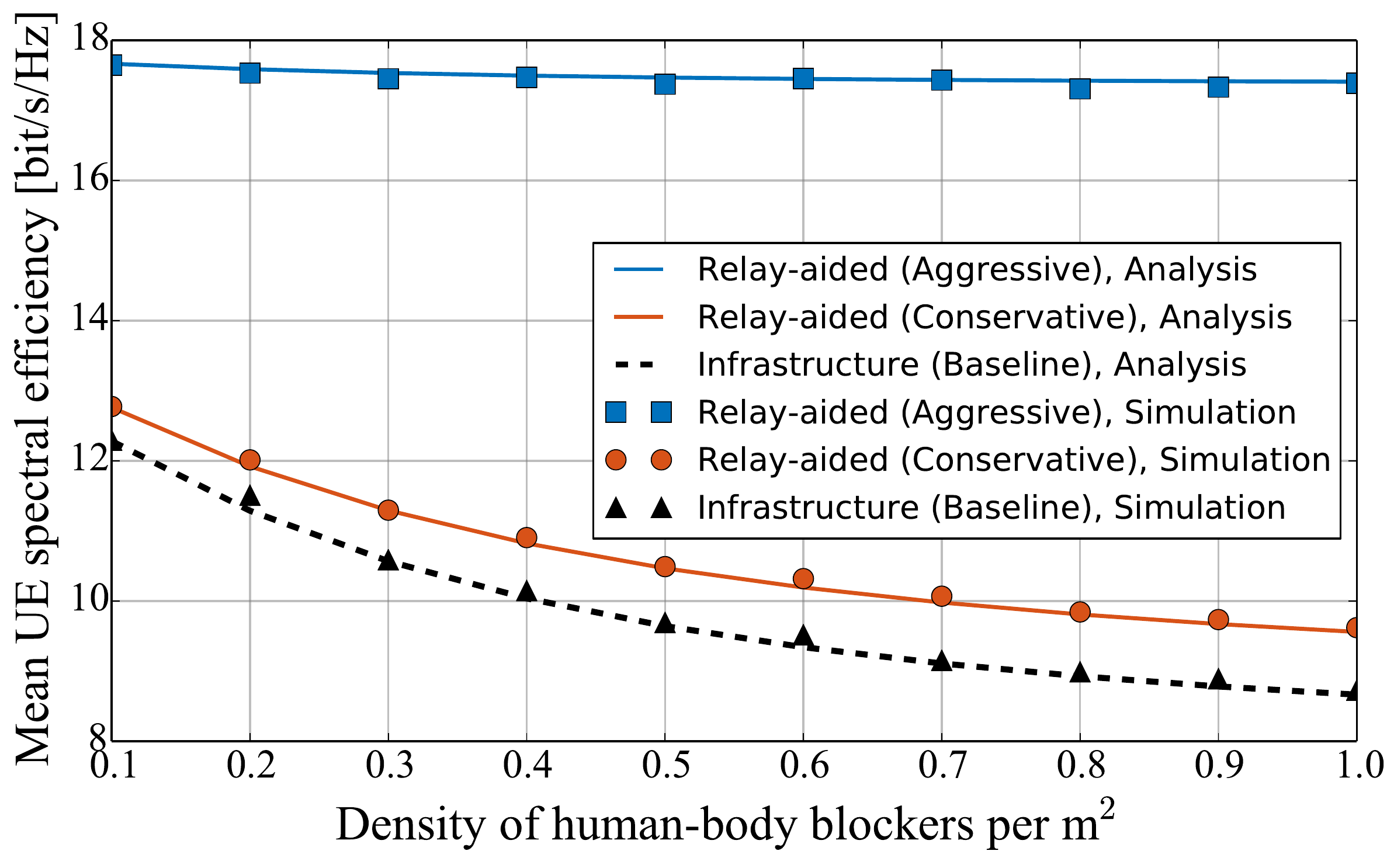}
    \vspace{-3mm}
    \caption{\textcolor{black}{Mean SE decreases with the growing density of human-body blockers. A close match between the analytical and simulation-based results is observed.}}
    \label{fig:plot1}
\end{figure}

We start with Fig.~\ref{fig:plot1} introducing the mean UE SE as a function of the density of human-body blockers on the sidewalks. We first observe that the mean SE decreases with the growing density of human-body blockers. For the Baseline strategy, the mean SE decreases from $12$\,bits/s/Hz for $0.1$\,humans/m$^2$ to $8$\,bits/s/Hz for $1.0$\,humans/m$^2$. Then, we notice that the gain of the considered relaying strategies grows for a higher density of blockers. In other words, more ``challenging'' environments unlock better gains. The SE with Aggressive strategy is notably higher than that in Conservative case, especially at higher densities of blockers: $17$\,bits/s/Hz vs. $9$\,bits/s/Hz for $1.0$\,human/m$^2$.

\begin{figure}[!t]
    \centering
    \includegraphics[width=0.9\columnwidth]{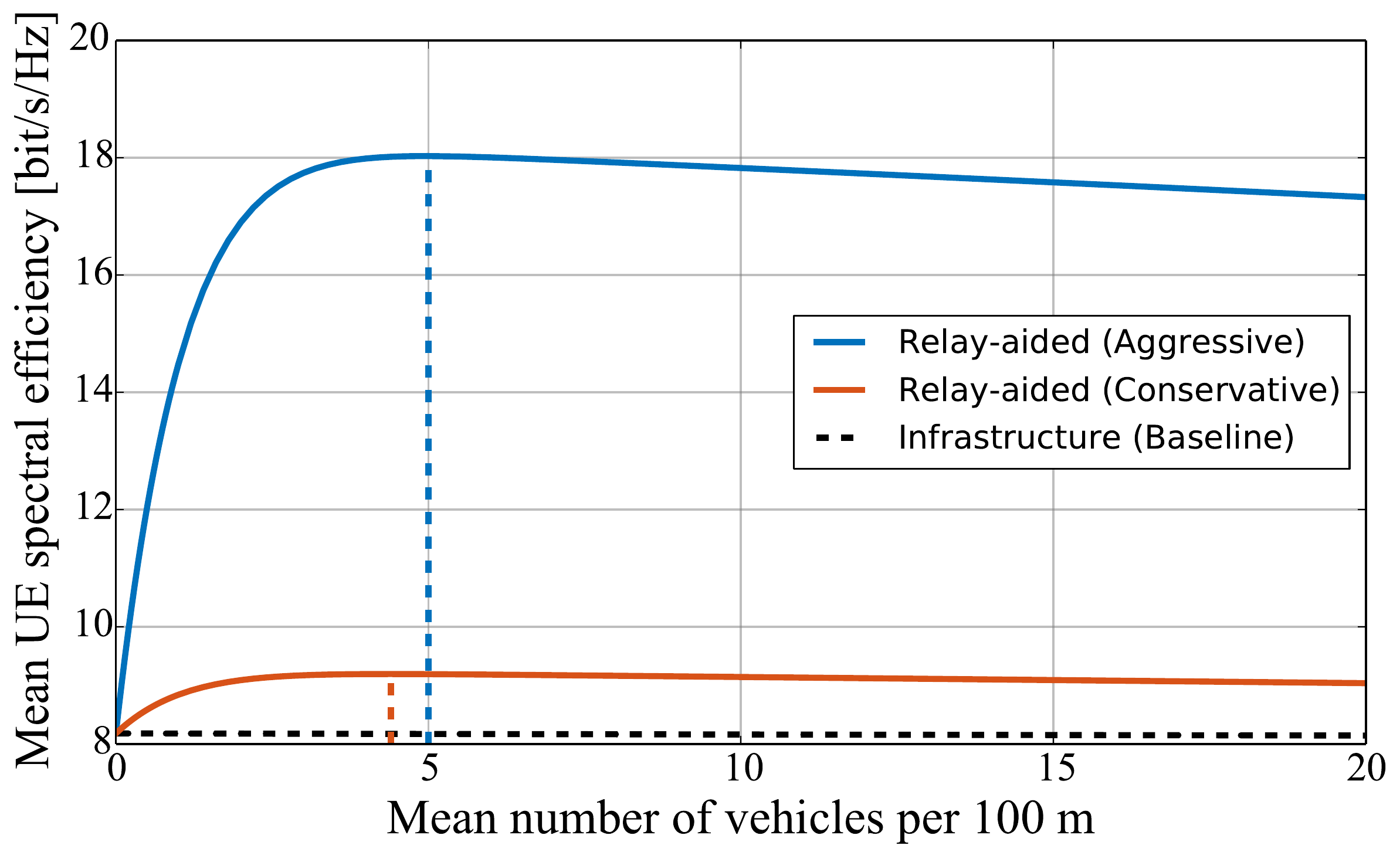}
    \vspace{-3mm}
    \caption{\textcolor{black}{High density of vehicles has a negative impact on the SE with Baseline strategy and a complex effect on the performance of relay-aided strategies.}}
    \label{fig:plot2}
\end{figure}

In Fig.~\ref{fig:plot1}, we also assess the accuracy of our derivations and the assumptions made by the system model. For this purpose, we relax three major analytical assumptions in our simulation framework: (i)~the pedestrians are not placed on the paths, but rather are uniformly distributed over the sidewalks; (ii)~vehicles are not centered on their lanes, but are randomly shifted within a lane, which models realistic city traffic; and (iii)~any of the vehicle's sides can block the mmWave signal, not only the one facing the communicating node. Fig.~\ref{fig:plot1} demonstrates a close match between the analytical and simulation-based results, even when the listed assumptions are relaxed. Similar correspondence between analysis and simulations is observed for other input parameters. Therefore, we rely solely on our analytical results in subsequent figures.

\subsubsection{The effect of vehicle density}
We proceed with Fig.~\ref{fig:plot2} that evaluates the mean UE SE as a function of the density of vehicles in the street. In this figure, we first note that only the Baseline scheme is always negatively affected by the growing density of vehicles. In contrast, the dependency is non-monotonic for relay-aided strategies. Particularly, the mean SE decreases at high vehicle densities as a result of vehicle-body blockage. The decrease in the mean SE at lower densities of vehicles is caused by the absence of vehicular relays. We can conclude that the highest gain of relaying is observed in deployments with the medium density of vehicles: over $9$\,bits/s/Hz with Conservative strategy and $18$\,bits/s/Hz with Aggressive strategy for $3$--$5$\,vehicles per~$100$\,m.

\subsubsection{The effect of COW coverage range}
\begin{figure}[!t]
    \centering
    \includegraphics[width=0.9\columnwidth]{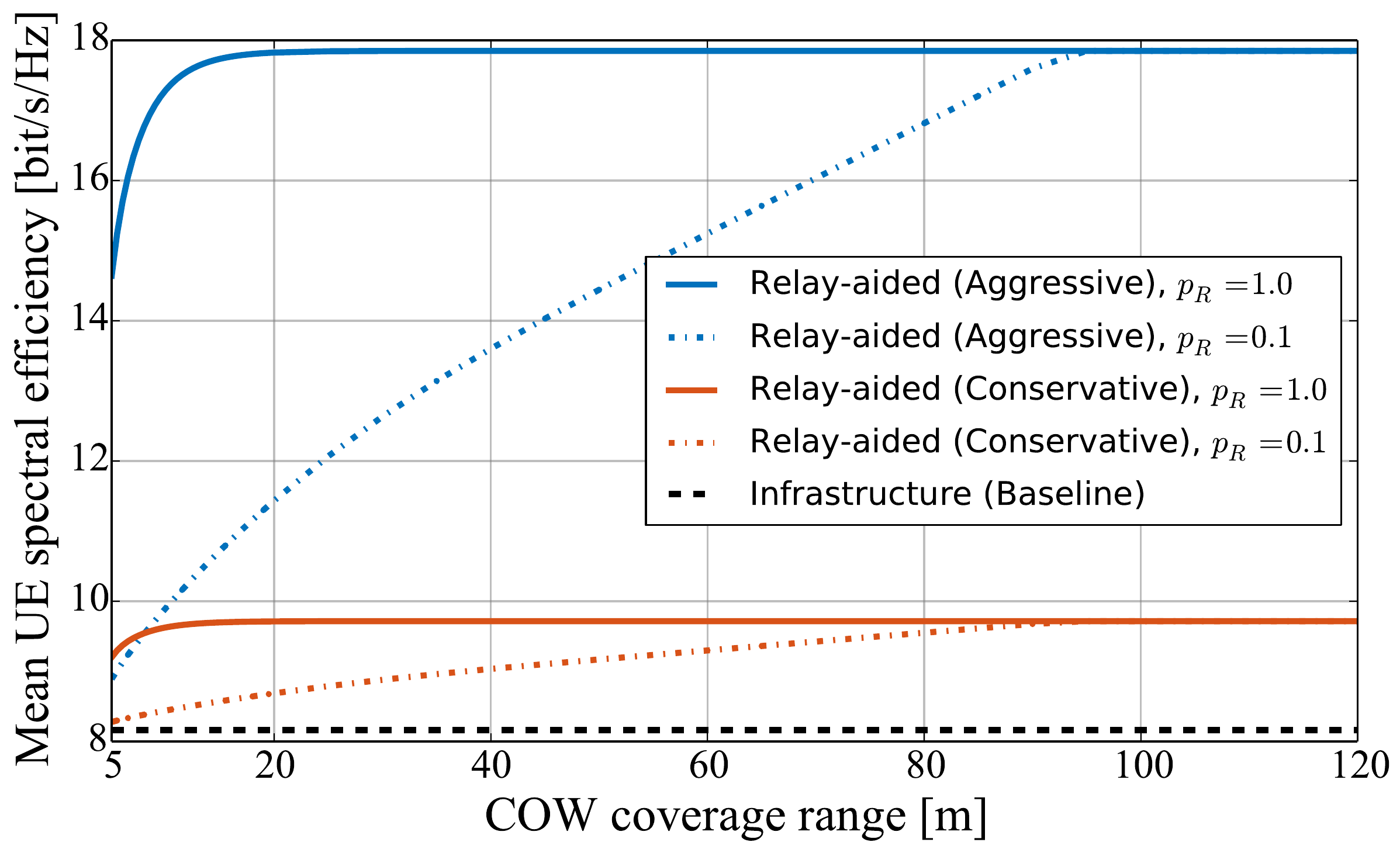}
     \vspace{-3mm}
    \caption{\textcolor{black}{Mean SE increases with COW coverage range and ceases to grow when the range becomes large enough so that $p_{\text{C}} \to 1$.}}
    \label{fig:plot3}
\end{figure}

We then analyze the impact of the COW coverage range on the mean UE SE in Fig.~\ref{fig:plot3}. Here, we first observe that the mean SE increases with the growing relay coverage range. Second, we notice that the SE ceases to grow when the range becomes large enough to almost guarantee a COW in the UE proximity. The value of COW range after which the mean SE stagnates heavily depends on the fraction of vehicles involved in relaying. Finally, we conclude that the Aggressive strategy notably outperforms the Conservative scheme: \textcolor{black}{the mean SE for the former with only $10$\% of COW vehicles is higher than the corresponding value for the latter when all $100$\% of~vehicles~act~as~COWs.}

\begin{figure}[!t]
    \centering
    \includegraphics[width=0.9\columnwidth]{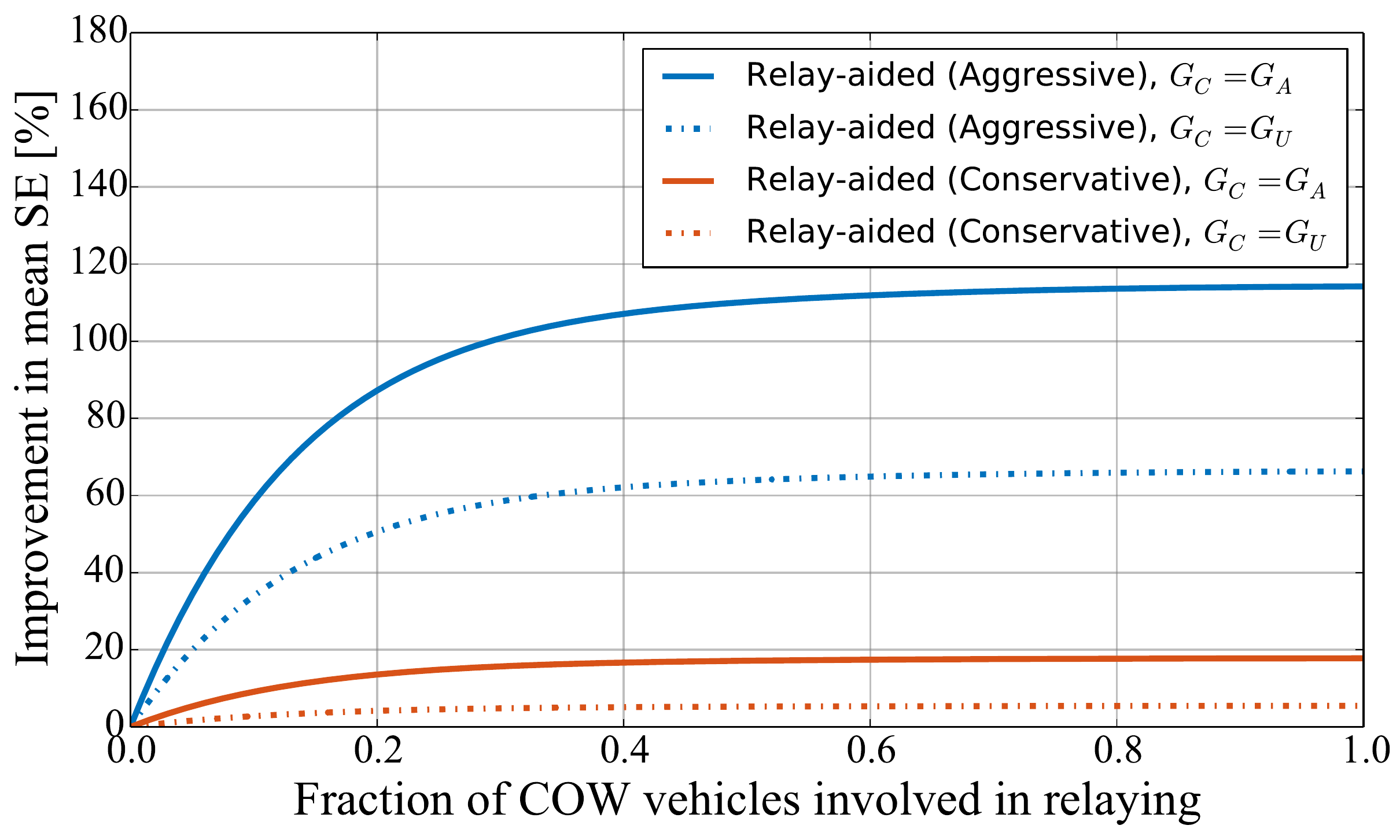}
    \vspace{-3mm}
    \caption{\textcolor{black}{Mean SE increases by $8$\%$-$$120$\% as $p_{\text{R}}$ grows. Even small fractions of COWs lead to notable performance gains with Aggressive strategy.}}
    \label{fig:plot4}
\end{figure}

\subsubsection{The effect of COW fraction}
Finally, we assess the overall increase in UE SE brought by COW relays in Fig.~\ref{fig:plot4}. Accordingly, the relative increase in the SE with respect to the Baseline strategy is presented as a function of the fraction of COW vehicles involved in relaying. Based on the obtained results, Conservative relays offer from $8$\% ($0.1$ vehicles act as COWs) to $12$\% (all vehicles act as COWs) increase in the mean UE SE. In the same conditions, Aggressive relays offer from $70$\% to $120$\% improvement. The gains with both strategies increase rapidly until $20$\% of vehicles are involved in relaying, whereas they cease growing after $40$\% of COW vehicles in the street. Therefore, there is no benefit in engaging more than $40$\% of vehicles in relaying. In contrast, the use of small fractions of COWs leads to notable performance gains.

\balance
\section{Conclusions}
\label{sec:conclusions}
In this paper, we proposed a performance analysis framework for the mmWave vehicular relaying in urban street layouts. We demonstrated that the performance gains brought by smart mmWave-based relaying are the most significant in vehicular deployments featured by dense human crowds on sidewalks and moderately-dense vehicular traffic. We also showed that even a small density of mmWave vehicular relays can lead to notable performance improvements provided that \textcolor{black}{intelligent} UEs can continuously select the best connectivity options \textcolor{black}{out of those offered by static mmWave APs and mmWave vehicular relays.}

\textcolor{black}{The developed framework and the presented results can further aid in identifying the setups where the use of mmWave vehicular relaying is especially beneficial, towards the adoption of mmWave-based mobile relays as part of 5G+ networks and their standardization in NR Rel.~17 and beyond.}

\section*{Acknowledgement}
\vspace{-1mm}
\label{sec:ack}
This work was supported by the Academy of Finland (projects WiFiUS, PRISMA, and RADIANT). R.~Heath was funded in part by the National Science Foundation under Grant No. ECCS-1711702 and by Honda R\&D Americas. 

\bibliographystyle{IEEEtran}
\bibliography{globecom_short}

\end{document}